\documentclass{elsart}

\usepackage{graphicx}
\usepackage{amssymb}

\begin{document}

\begin{frontmatter}

% Title, authors and addresses

% use the thanksref command within \title, \author or \address for footnotes;
% use the corauthref command within \author for corresponding author footnotes;
% use the ead command for the email address,
% and the form \ead[url] for the home page:

\title{The Transient Growth of Ammonium Chloride Dendrites}
\author{Andrew Dougherty\corauthref{cor1}} and
\corauth[cor1]{Corresponding author.}
\ead{doughera@lafayette.edu}
\ead[url]{http://www.lafayette.edu/$\sim$doughera}
\author{Thomas Nunnally}
\address{Dept.~of Physics, Lafayette College, Easton PA 18042}

\begin{abstract}
We report measurements of the initial growth and subsequent
transient response of dendritic crystals of ammonium chloride grown
from supersaturated aqueous solution.  Starting from a small, nearly
spherical seed held in unstable equilibrium, we lower the temperature
to initiate growth.  The growth speed and tip radius approach the same
steady state values independent of initial seed size.  We then explore
the response of the growing dendrite to changes in temperature.
The crystal adjusts quickly and smoothly to the new growth conditions,
maintaining an approximately constant value of $v \rho ^2$ throughout.
Dissolving dendrites, on the other hand,
are not characterized by the same value of
$v \rho ^2$.

% $Revision: 1.23 $
\end{abstract}

\begin{keyword}
% keywords here, in the form: keyword \sep keyword
A1. Dendrites \sep A2. Growth from solution \sep A1. Crystal morphology
% PACS codes here, in the form: \PACS code \sep code
\PACS 68.70.+w \sep 81.10.Dn \sep 64.70.Dv
\end{keyword}
\end{frontmatter}

\section{Introduction}

Dendritic crystals are commonly observed when a non-faceted material
grows from a supercooled melt or supersaturated solution.  Metals and
metal alloys are perhaps the most technologically important applications,
but dendrites are also observed in some transparent organic compounds
and in some salt solutions.  For broad overviews with additional
references, see Refs.~\cite{Trivedi00,Glicksman-rev}.  Mathematically,
dendritic crystal growth is a challenging moving boundary problem, and
is one of the canonical problems of pattern formation in non-linear,
non-equilibrium systems \cite{Langer80}.

One steady-state solution for the growth of a crystal is an approximately
parabolic dendrite with radius of curvature $\rho$ propagating at
constant speed $v$ determined by the material properties and growth
conditions \cite{Kessler88a,McFadden2000}.  Real dendrites, however,
are not simple steady-state crystals.  The smooth shape is unstable to
sidebranching perturbations, resulting in a complex time-dependent shape
where a compete specification of the state at any time could depend on
the previous growth history \cite{Schaefer78,Glicksman99c}.

Numerical simulations, particularly phase field models, have
made considerable progress in following this time-dependent
evolution and extracting the steady state results.  For reviews, see
Refs.~\cite{Karma2001,Boettinger2002}.  Careful experimental measurements
are still needed to determine whether those models indeed contain all
of the relevant physics, and how well they apply to noisy irregular
dendrites growing under actual experimental conditions.

In this work, we examine the role of transients in dendritic crystal
growth.  In particular, we look at the initial emergence and growth of
a dendritic crystal from a well-characterized nearly spherical seed.
We then monitor the changes in the crystal when it is subjected to
carefully-controlled changes in growth conditions.
The experimental protocol is designed to be readily reproducible in
numerical simulations.

\section{Experiments}

The experiments were performed with aqueous solutions of ammonium chloride
with approximately 38\% NH$_4$Cl by weight.
% Book #4, pg. 055.
The saturation temperature was approximately $71^\circ$C.  The solution
was placed in a $40 \times 10 \times 2$ mm glass spectrophotometer cell
sealed with a Teflon stopper held in place by epoxy.
The cell was mounted in a massive copper block, surrounded by an outer
aluminum block, and placed on a microscope enclosed in an insulating box.

The temperature of the outer aluminum block was controlled to
approximately $\pm 1^\circ$C.  The temperature of the inner copper block
was controlled directly by computer, allowing complete programmatic
control over the temperature during the course of a run.  The temperature
of the sample was stable to within approximately
$\pm 5 \times 10^{-4\ \circ}$C.

Images were obtained from a charged coupled device (CCD) camera attached
to the microscope and acquired directly into the computer via a Data
Translation DT3155 frame grabber with a resolution of $640 \times
480$ pixels.  The ultimate resolution of the images was $0.63 \pm 0.01
\mu$m/pixel.  The interface position was determined in the same manner
as in Refs.~\cite{Dougherty88,Dougherty2005}.

To obtain a single crystal, the solution was heated to dissolve all the
NH$_4$Cl, stirred to eliminate concentration gradients, and then cooled to
initiate growth.  Many crystals would nucleate, but an automated process
was set up to acquire images and slowly adjust the temperature until only
a single isolated crystal remained.  The temperature was then continually
adjusted until this crystal was the desired size.  Although such a
spherical crystal is in a state of unstable equilibrium, we found it
possible to maintain it indefinitely, provided we continually monitored
the size and adjusted the temperature accordingly.  This isolated,
nearly spherical, crystal was allowed to stabilize for several days.

For an experimental run, the temperature was first held constant for 500
seconds, and then lowered steadily.  The typical drop was $1.2^\circ$C
at a rate of $1^\circ$C$/600$s.  The time when the temperature started
dropping was taken as $t = 0$.  During the run, images were taken at
1-second intervals.

\section{Initial Growth from a Nearly Spherical Seed}

\begin{figure}
\includegraphics[keepaspectratio=true,
width=\columnwidth]{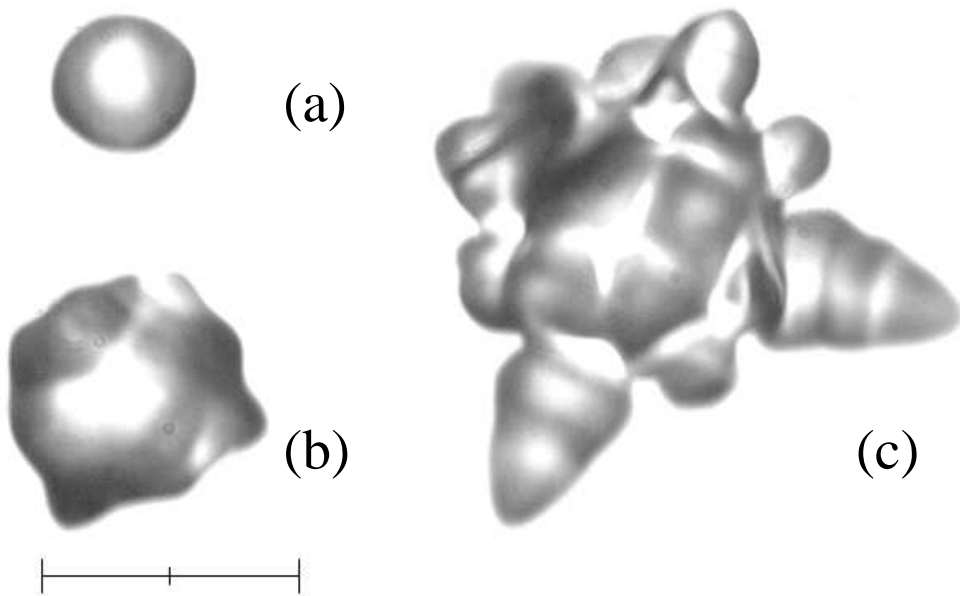}
\caption{Initial stages of growth from a nearly
spherical seed at 0s (a), 350s (b), and 850s (c)
after the temperature began to drop.  The scale bar is 100$\mu$m long.}
\label{composite}
\end{figure}

The initial stages of a growing crystal of NH$_4$Cl are shown in
Fig.~\ref{composite}.
Initially, the crystal remains approximately spherical as it grows.
As it gets larger, however, the smooth surface becomes unstable and
precursors to dendritic tips emerge.  Since NH$_4$Cl has cubic symmetry,
six dendrites would normally be expected in the [100] directions---four
in the plane of the image, and two perpendicular to that plane.  (The
asymmetry in Fig.~\ref{composite} is due to the imperfect orientation of
the crystal.)  In addition, there are smaller protuberances at angles
between the main dendrite arms corresponding to less-favored growth
directions \cite{Chan76,Chan78}.

We model the initial stages of the growth as a quasi-static, spherically
symmetric, and diffusion-limited process.  Under those constraints,
the radius $R$ of the crystal is given by

\begin{equation}
\frac{d R}{d t} = \frac{D}{R} \left ( \Delta - \frac{ 2 d_0}{R} \right ) ,
\label{eq:dRdt}
\end{equation}

where $D$ is the diffusion coefficient for NH$_4$Cl in aqueous solution,
$\Delta$ is the dimensionless supersaturation, and $d_0$ is the
capillary length, which incorporates the surface energy \cite{Langer80}.
The diffusion coefficient for lower concentrations and temperatures was
measured by Lutz and Mendenhall \cite{Lutz2000} and by Hall, Wishaw,
and Stokes \cite{Hall53}.  Extrapolating those published values,
we estimate $D = 2500 \mu$m$^2/$s.  For this experiment, where the
crystal rests against the bottom plate of the growth cell, we estimate
the effective diffusion constant to be $0.71 D$ \cite{Dougherty88}.
Lastly, $\Delta$ was assumed to be linearly
related to the temperature $T$ and the bulk equilibrium temperature
$T_{eq}$ by

\begin{equation}
\Delta = \frac{d\Delta}{dT} ( T_{eq} - T ) .
\label{eq:Delta}
\end{equation}

The coefficient $d\Delta / dT$ and the capillary
length $d_0$ were determined by fits to Eq.~\ref{eq:dRdt}.

\begin{figure}
\includegraphics[width=\columnwidth]{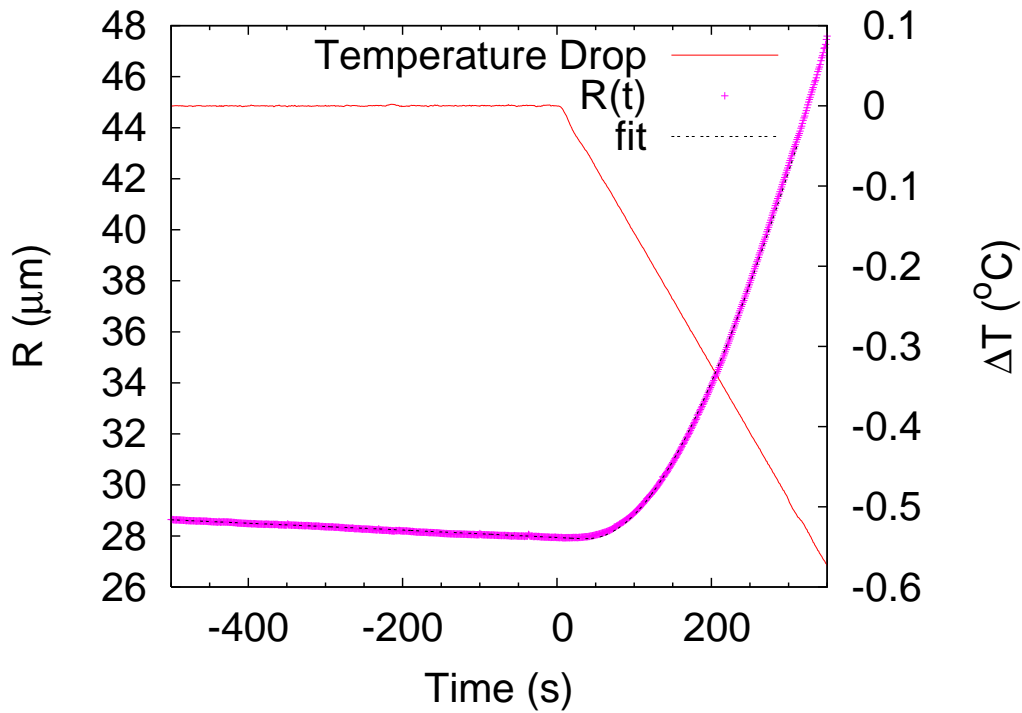}
\caption{Radius of a growing crystal (+) along with a fit
to Eq.~\ref{eq:dRdt} (dashed line).
The temperature profile for this run (solid line) is shown as well.}
\label{Rfit}
\end{figure}

The radius as a function of time for the crystal in
Fig.~\ref{composite} is shown in Fig.~\ref{Rfit}, along with a fit to
Eq.~\ref{eq:dRdt}.
The fit only includes times prior to that corresponding to
Fig.~\ref{composite}(b), since the crystal shape
becomes significantly distorted away from spherical after those times.
From the fit shown in Fig.~\ref{Rfit}, we estimate $d\Delta / dT =
0.0050 \pm 0.0005 /^\circ$C.  We also used $d_0 = 3\times 10^{-4} \mu$m,
but the fit is not very sensitive to that value, since the second term
in Eq.~\ref{eq:dRdt} is small for the large crystals considered here.

\section{Emergence of Dendrites}

To study the transition from initial sphere to fully-developed dendrites,
we considered two runs with identical temperature profiles; one starting
from a relatively small seed, and the
second from a much larger seed.  As
above, we initially held the temperature constant for 500 seconds, and
then lowered the temperature $1.2^\circ$C at a rate of
$1^\circ$C$/600$s.

\begin{figure}
\includegraphics[width=\columnwidth]{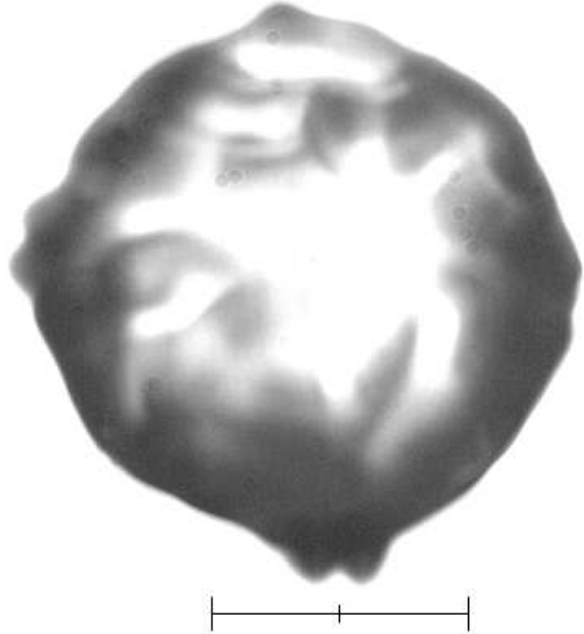}
\caption{Early stage of dendritic growth
from a large seed.  This image was taken
550s after the temperature began to drop. The scale bar is 100$\mu$m
long.}
\label{large_seed}
\end{figure}

The initial growth of the small seed was shown above in Fig.~\ref{composite}.
The large seed is shown in Fig.~\ref{large_seed} at a time of 550s
after initiating growth.
Near the bottom of the image, instead of a single dendrite arm, there are
two nearly equal arms competing.  Eventually, the one on the right won
out, and was followed for the remainder of the run.

\begin{figure}
\includegraphics[width=\columnwidth]{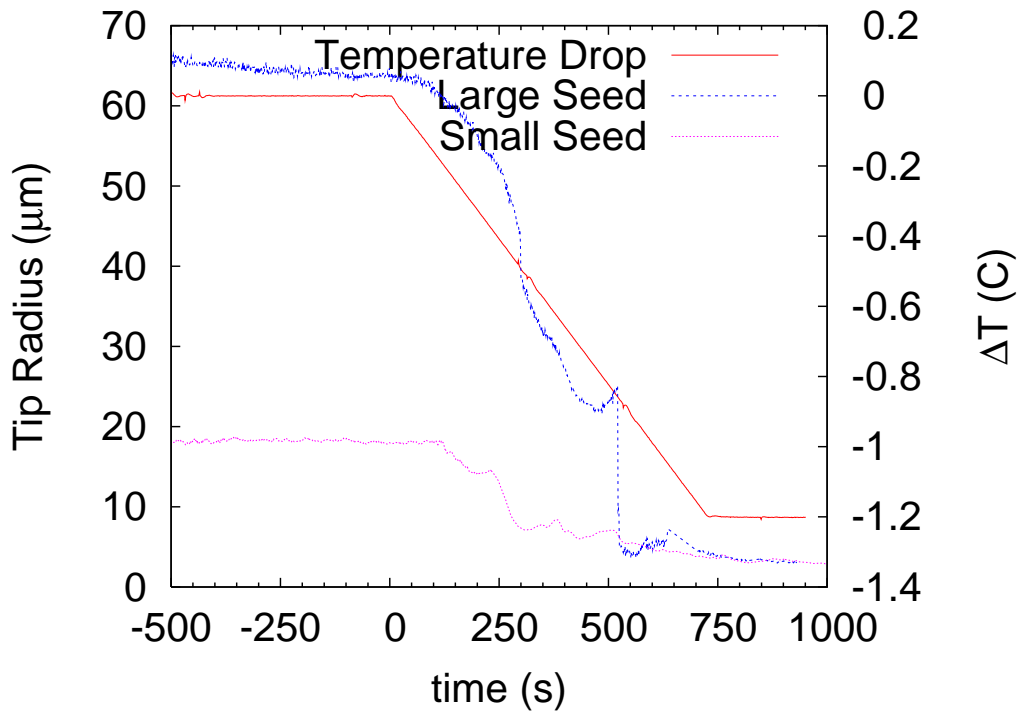}
\caption{Initial evolution of the radius of curvature for
large (dashed) and small (dotted) seeds.
Both approach the same steady state value.
The temperature profile (solid line) is shown as well.}
\label{init_rhos}
\end{figure}

The evolution of the tip radius of curvature for the two crystals is
shown in Fig.~\ref{init_rhos}.
The curvature was measured by fitting a parabola with a
fourth-order correction as in Ref.~\cite{Dougherty2005}:

\begin{equation}
z = z_{tip} + \frac{(x-x_{tip})^2}{2\rho} -
A_4 \frac{(x-x_{tip})^4}{\rho^3} ~,
\label{eq:fourth}
\end{equation}

where the growth direction of the dendrite is taken to define the
$-z$ direction, $(x_{tip}, z_{tip})$ is the location of the tip,
$\rho$ is the radius of curvature
at the tip, and $A_4$ is an orientation-dependent fourth-order
correction.  For fully-established steady-state dendrites, the most
robust results were obtained if data were taken up to a distance of
$z_{max} = 6 \rho$ behind the tip.  For incipient dendrites, such as
those in Fig.~\ref{large_seed}, only data a small distance of $z_{max} =
0.2 \rho$ could be used.

For the large seed, there is a sharp drop near 500s; this corresponds
to the time shortly before that shown in Fig.~\ref{large_seed}, when
the instability on the bottom is beginning to develop into two distinct
dendrites, and the fitting program switches from fitting an envelope
including both emerging tips to a fit including just a single tip.
In such cases, the ``best'' fit is not well-defined, but the graph
does accurately reflect the emergence of dendrites.  Both large and small
seeds approach the same steady-state value to within the experimental
uncertainty.  For the time interval after the temperature reached its
steady-state value, 750--1000s, the average radius for the dendrite
grown from either the large or the small seed is $3.3 \pm 0.2 \mu$m.
% Notebook #4, pg. 080.

\begin{figure}
\includegraphics[width=\columnwidth]{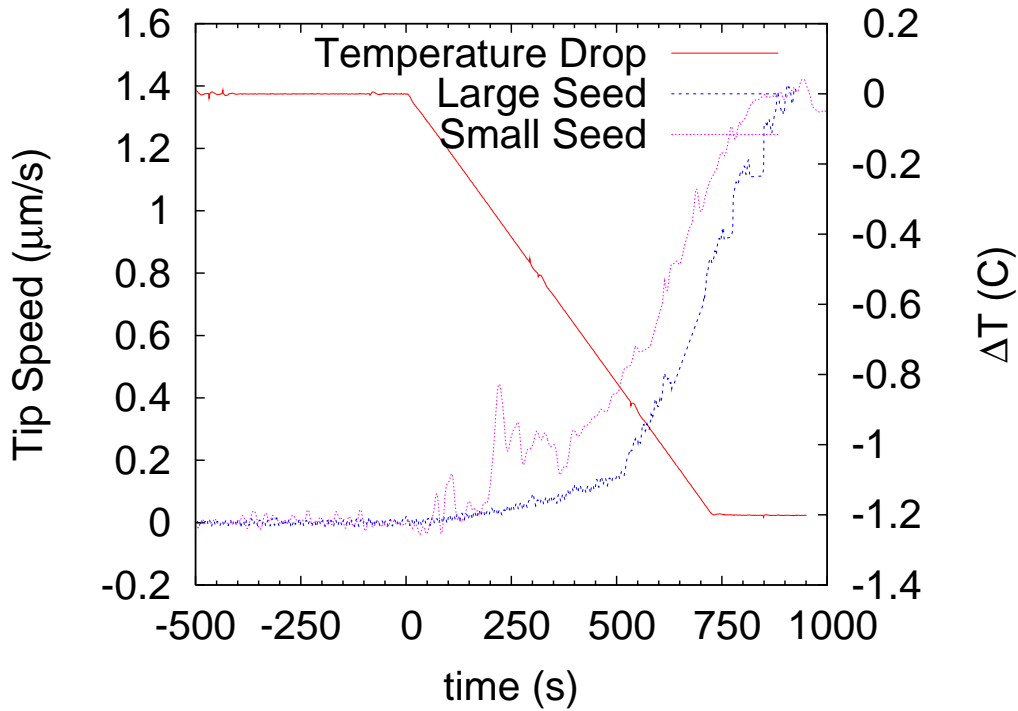}
\caption{Initial evolution of the tip speeds for
large (dashed) and small (dotted) seeds.
Both approach the same steady state value.}
\label{init_speeds}
\end{figure}

The tip speed was determined simply by measuring the displacement
of $(x_{tip}, z_{tip})$ over time.  The initial evolution of the tip
speed is shown in Fig.~\ref{init_speeds}.  The fluctuations for the
small seed near 250s correspond to times shortly before that shown
in Fig.~\ref{composite}(b), when the initial tips are just becoming
discernible, and the fit is not well-defined.  Again, both approach the
same steady-state value of $1.3 \pm 0.1 \mu$m/s, though the large seed
does so slightly more slowly, perhaps due to the competition between
adjacent tips visible at the bottom of Fig.~\ref{large_seed}.

This behavior is somewhat different from that observed in the simulations
of Steinbach, Diepers, and Beckermann \cite{Steinbach2005}, where
the approach to steady state differed depending on initial seed
size, but that difference is likely due to different initial conditions.
In Ref.~\cite{Steinbach2005}, the authors used initial conditions
appropriate for the Isothermal Dendritic Growth Experiments (IDGE) of
Glicksman and co-workers \cite{Glicksman94a,Glicksman99a,Glicksman99b}.
In those experiments, the growth chamber was held at a constant
supercooling, and growth was initiated through a capillary tube, or
``stinger''.  The initial crystals that grew from the end of the stinger
entered a region where the temperature was very nearly uniform, and the
steady-state temperature field had not yet been established.  Steinbach,
Diepers, and Beckermann modeled this as growth from a finite-sized seed
placed in a region of originally uniform supercooling.  Their simulations
showed that the initial transient approach to steady state depended on
the size of the initial seed.

In this work, the initial seed is in equilibrium with the saturated
solution.  The temperature is then lowered, initiating growth.  The fit in
Fig.~\ref{Rfit} demonstrates that the present system is reasonably well
described as being in quasi-static equilibrium throughout the initial growth
process.

The possible significance of initial seed size was also examined by
Pines, Chait, and Zlatkowski \cite{Pines97}.  They considered thermal
growth at very low undercooling, and concluded that there ought to be
corrections on the order of $\rho/R_0$, where $\rho$ is the steady
state tip radius, and $R_0$ is the radius of the initial sphere.
That ratio is 0.04 and 0.12 for the large and small seeds,
respectively, which corresponds to a predicted difference of
approximately $0.26 \mu$m, comparable to the uncertainties in the
measurements.
% Book #4, pg. 080.

\section{Response to Transients}

To study the response of growing dendrites to transients, the crystals
were subjected to a series of temperature changes, and the tip radius
$\rho$ and speed $v$ were monitored throughout.
After the initial transient, the
temperature was held constant for a long time, and then the temperature
was changed at approximately $1^\circ$C$/200$s, which is as quickly as the
apparatus could respond.

\begin{figure}
\includegraphics[width=\columnwidth]{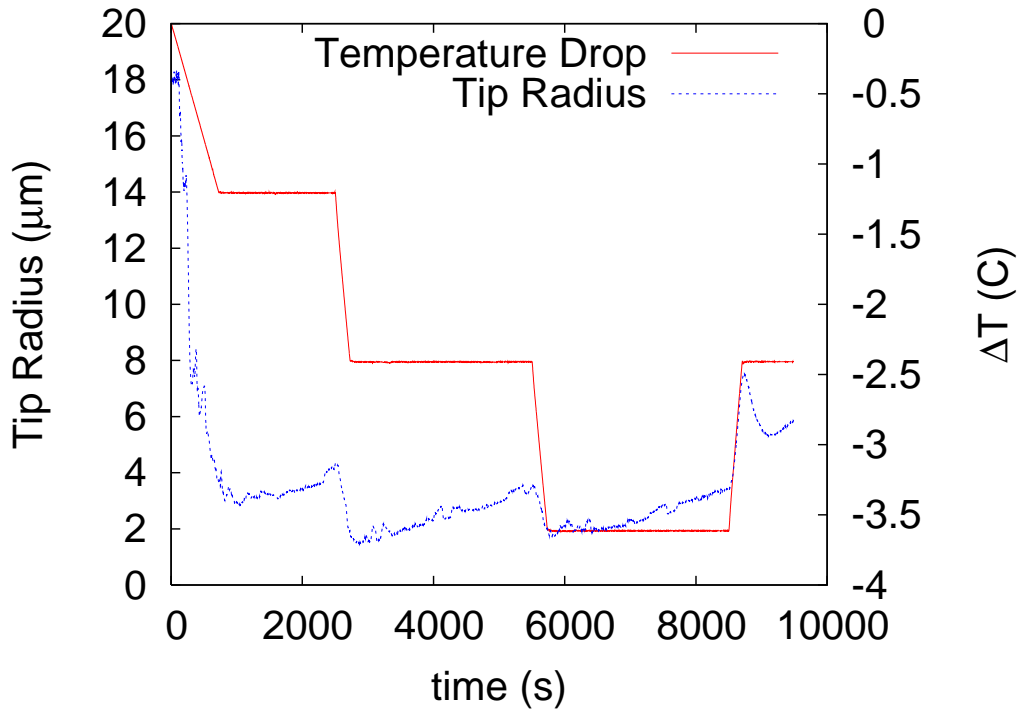}
\caption{Response of the tip radius (dashed line) to transients.  The temperature
profile (solid line) is shown as well.  The temperature was changed at 2500, 5500,
and 8500s.}
\label{rhovst}
\end{figure}

The evolution of the tip radius is shown in Fig.~\ref{rhovst}.
During each segment at constant temperature, the growing
crystal slowly depletes the growth cell, so the tip radius gradually
increases and the tip speed gradually decreases.  (Similar finite size
effects were also seen in the IDGE results at very low supercoolings
\cite{Glicksman94a,Glicksman99a,Glicksman99b,Pines96,Sekerka97}.)
Each subsequent lowering of the temperature brings on a fresh burst
of growth.

The primary cause of the scatter in the data is the slight change in
focus as the crystal grows across the screen.  Since the crystal does
not grow exactly in the horizontal plane, the focus changes slightly
during the course of the run, and has to be periodically adjusted.
This leads to small fluctuations in the measured value for $\rho$.

\begin{figure}
\includegraphics[keepaspectratio=true, width=0.8\columnwidth]{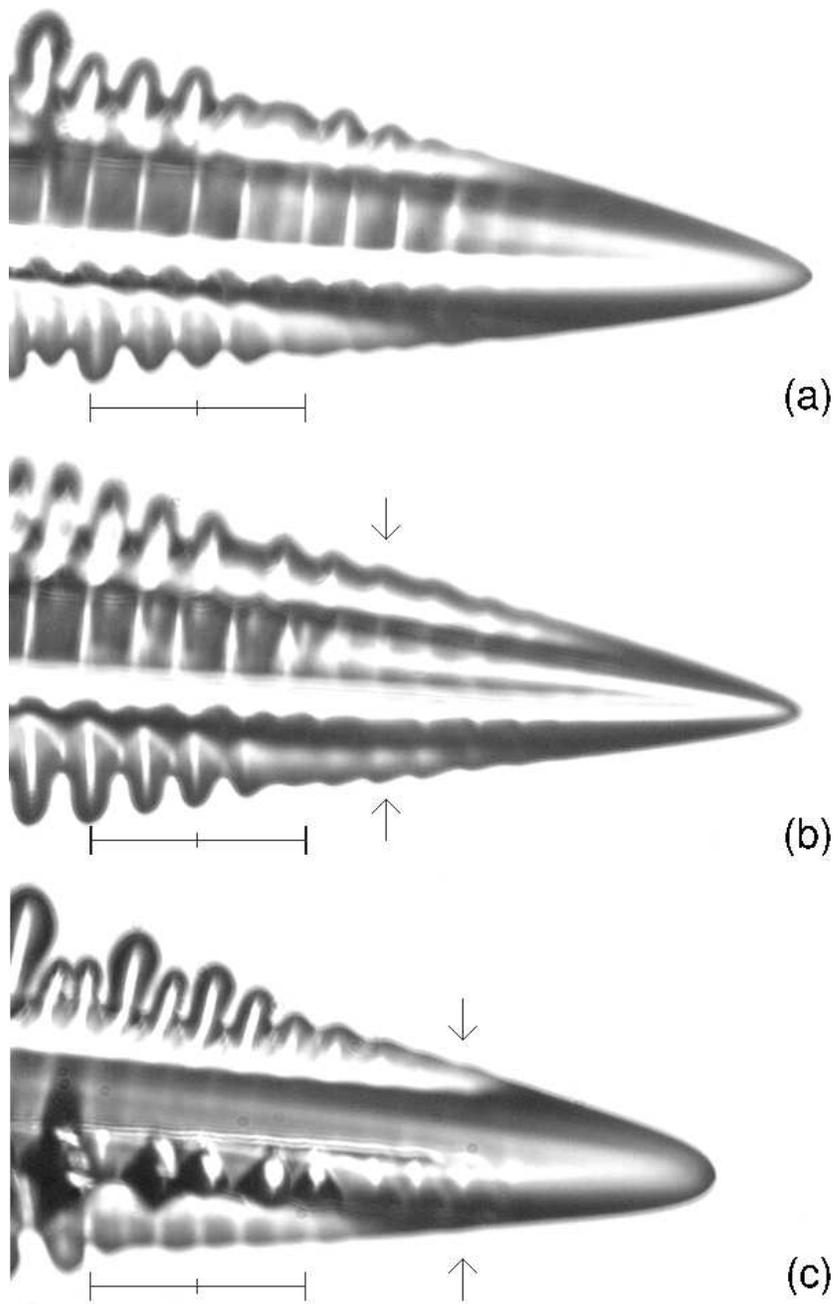}
\caption{Images of crystal at (a) $t=2325s$, shortly
before the temperature drop at $t=2500s$, (b) $t=2673s$, shortly
after the temperature drop, and (c) at $t=8720s$, when the speed dropped
to zero after the temperature increase.  The scale bars are 100$\mu$m
long.  The arrows in (b) and (c) show the tip location at the times
when the transients started.}
\label{before_and_after}
\end{figure}

The changes in crystal morphology shortly before and after
transients are shown in Fig.~\ref{before_and_after}.  A steady state dendrite
at $t = 2325$s is shown in Fig.~\ref{before_and_after}(a).
There are fairly regular sets of sidebranches growing on each side.
Fig.~\ref{before_and_after}(b) shows the crystal at $t=2673$s, after
the temperature began to drop at 2500s.  The
arrows show the location of the tip at $t=2500$s.
Because of the decrease in temperature, the tip speed
is higher and the tip radius is smaller.  Finally,
Fig.~\ref{before_and_after}(c) shows the crystal at $t=8720$s, when
the speed dropped to zero after the temperature was increased at
8500s.
The arrows show the location of the tip at $t=8500$s.

Overall, the transients in this case are slow enough that the crystal is able
to adjust relatively smoothly to the change, without the generation of
large sets of sidebranches.   Other experiments with more rapid
transients have seen large sets of sidebranches generated
\cite{Cummins93,Borzsonyi2000,Couder2005,Bilgram2006}.  We also saw no
evidence of any
transition to doublons, as has been observed in xenon dendrites
\cite{Bilgram2004a,Bilgram2005}.

During the transitions, the tip curvature is not
precisely defined, since the fit to Eq.~\ref{eq:fourth} attempts to
include portions of the crystal grown under different conditions, but
the distortion is not too significant, as seen in Fig.~\ref{rhovst}.
At $t=8500$s, the temperature was raised back up so that the crystal
started to dissolve.  While dissolving, the crystal
shape deviates significantly from that given by
Eq.~\ref{eq:fourth}, though the measured tip curvatures are still a
reasonable representation of the tip size.

\begin{figure}
\includegraphics[width=\columnwidth]{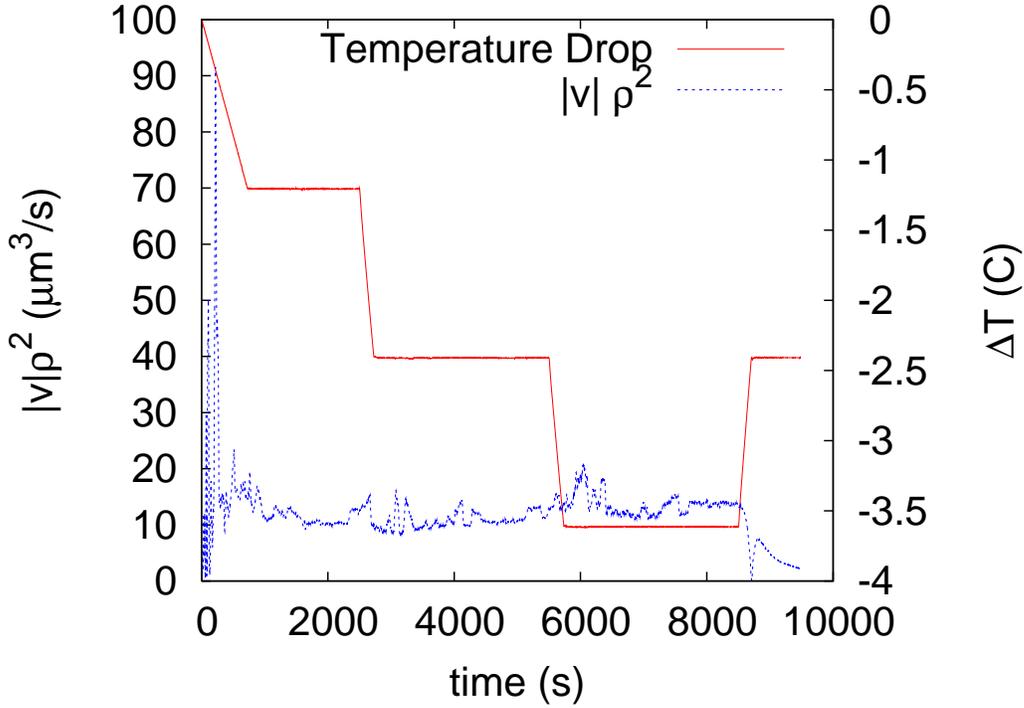}
\caption{ Evolution of $|v| \rho^2$ (dashed line) in response to
transients.
The large fluctuations at the beginning of the run correspond to the
initial emerging dendrites.}
\label{vr2vst}
\end{figure}

Finally, throughout the experiment, the quantity $|v| \rho^2$ remains
approximately constant during growth, as shown in Fig.~\ref{vr2vst}, with an
average value of $12 \pm 2 \mu$m$^2/$s.
There are no significant changes during the transients.  The large
fluctuations near the beginning are due to the initial growth of the
instability, exacerbated by the ambiguities inherent in determining
the tip radius and position of incipient dendrites, such as those in
Fig.~\ref{composite}(b).  Around $t=6000$s, the crystal reached the edge of
the viewing area, and the growth cell was moved to bring it
back into view.  This apparently caused a small bit of motion inside
the cell, leading to a momentary increase in speed.  After $t=8720$s,
the crystal began dissolving, and the velocity became negative.  During the
dissolution, the quantity $|v| \rho^2$ was no longer constant.

These results are consistent with the
findings of Steinbach \textit{et al.}, who observed significant
transients in their simulations, but estimated that the time for
transient responses is of the order of $\rho/2 v$ \cite{Steinbach2005},
which is of the order of 1.0s in the present experiments.

\section{Conclusions}

We have performed experiments on the growth of NH$_4$Cl dendrites in
which the growth conditions have been carefully controlled.  We find that
the approach to steady-state growth is rather robust; the same tip
radius and speed are obtained independent of initial seed size.
We also find that the crystal smoothly adjusts to transients that
are reasonably slow compared to the time scale $\rho / v$.
Future experiments will focus on faster transients and on the response
of the sidebranching structure to those transients.

% Put \label in argument of \section for cross-referencing
%\section{\label{}}
% \subsection{}
% \subsubsection{}

% Specify following sections are appendices. Use \appendix* if there
% only one appendix.
%\appendix
%\section{}

% If you have acknowledgments, this puts in the proper section head.
%\begin{acknowledgments}
% put your acknowledgments here.
%\end{acknowledgments}

% Create the reference section using BibTeX:
\bibliography{init_trans}
\bibliographystyle{elsart-num}

\end{document}